# Temperature-Dependent Polarized Raman Spectra of $CaFe_2O_4$

N. Kolev and M. N. Iliev

*Texas Center for Superconductivity and Advanced Materials (TcSAM)*
*University of Houston, Houston, TX 77204-5002, USA*

V. N. Popov

*Faculty of Physics, University of Sofia, 1164 Sofia, Bulgaria*

M. Gospodinov

*Institute of Solid State Physics, Bulgarian Academy of Sciences, 1184 Sofia, Bulgaria*

The Raman spectra of $CaFe_2O_4$ were measured with several exact scattering configurations between 20 and 520 K and the symmetry of all observed Raman lines was determined. The $A_g$ and $B_{2g}$ lines were assigned to definite phonon modes by comparison to the results of lattice dynamical calculations. No anomaly of phonon parameters was observed near the magnetic ordering temperature $T_N = 160$ K.

$CaFe_2O_4$ crystallizes in the orthorhombic *Pnma* structure (#62, *Z*=4) made up of distorted $FeO_6$ octahedra sharing edges and corners, and eight-fold coordinated calcium atoms (Fig1). Below $T_N = 160$ K the magnetic moments of $Fe^{3+}$ (S=5/2) order antiferromagnetically. The structure, magnetic and thermodynamic properties of $CaFe_2O_4$ have been studied in number of experimental works [1-9], however, there are scarce data on the electronic, transport and optical properties. In particular, neither experimental, nor theoretical results on the phonons in $CaFe_2O_4$ are available so far. Such data are of definite interest for better understanding of the properties of $CaFe_2O_4$ and a wide variety of materials with $CaFe_2O_4$-type structure.

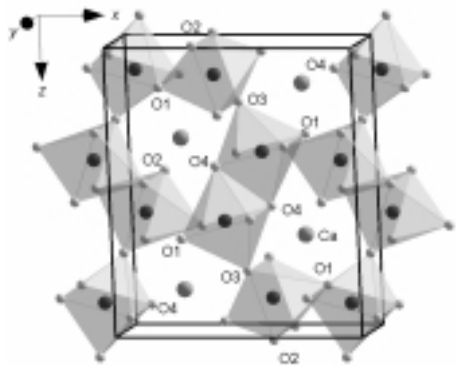

Fig.1 Elementary cell of CaFe2O4.

In this work we present and discuss the polarized Raman spectra of $CaFe_2O_4$. The observed Raman lines are assigned to definite phonon modes on the basis of their symmetry and comparison to the frequencies predicted by lattice dynamical calculations (LDC). Between 20 and 520 K the Raman lines show the standard temperature shift expected for anharmonic scattering. No anomaly of phonon parameters is observed near the magnetic ordering temperature $T_N = 160$ K.

Single crystals $CaFe_2O_4$ were grown by the high temperature solution growth method using $CaCl_2$ flux. The flux was mixed with stoichiometric amount of $CaO + Fe_2O_3$ powder in a 20:1 ratio and annealed in a closed platinum crucible at $1350^{\circ}C$ for 24h in oxygen. Then the temperature was decreased down to $1000^{\circ}C$ at a rate of $2^{\circ}C/h$. The flux was soluble in water and well-shaped whisker-like prisms of typical size $20 \times 0.2 \times 0.2$ mm$^3$ were removed from the bottom of the crucible. The elemental content was confirmed by SEM.

The Raman spectra were measured under microscope using 514.5 nm Ar$^+$ laser line. For temperature-dependent measurements the sample was mounted in an optical cryostat where the temperature could be varied between 10 and 520 K. The polarization dependence of the Raman spectra has confirmed that as a rule the short crystal edges are along the crystallographic (101) and (10$\underline{1}$) directions, whereas the long edges are always along the (010) direction. Further, the standard notations *x*, *y*, *z*, *x'*, and *z'* will be used for the (100), (010), (001), (101) and (10$\underline{1}$) directions, respectively.

In the orthorhombic unit cell of $CaFe_2O_4$ all atoms occupy *4c* sites of $C_s^{xz}$ symmetry and each participates in 12 ($2A_g + B_{1g} + 2B_{2g} + B_{3g} + A_u + 2B_{1u} + B_{2u} + 2B_{3u}$) Γ-point phonon modes. Out of in total 84 modes 42 ($14A_g + 7B_{1g} + 14B_{2g} + 7B_{3g}$) are Raman active. The non-zero elements of the Raman tensors for the $A_g$ ($\alpha_{xx}, \alpha_{yy}, \alpha_{zz}$), $B_{1g}$ ($\alpha_{xy}$), $B_{2g}$ ($\alpha_{xz}$), and $B_{3g}$ ($\alpha_{yz}$) modes determine the polarization selection rules. The $A_g$ modes are allowed with parallel *xx*, *yy*, *zz*, *x'x'*, *z'z'* and forbidden with crossed *xy*, *xz*, and *yz* scattering configurations. The first and second letters in these notations denote the polarizations of incident and scattered light, respectively. The $B_{1g}$, $B_{2g}$, and $B_{3g}$ modes are allowed with crossed *xy*, *xz* and *yz* but forbidden with parallel *xx*, *yy* and *zz* configurations, respectively. They are however allowed in parallel polarizations (e.g. *x'x'* for a $B_{2g}$ mode) rotated at 45 degree with respect to the main crystallographic directions.

The Raman spectra of $CaFe_2O_4$ as obtained at room temperature with several exact scattering configurations are shown in Fig.2. The lines of $A_g$ or $B_{2g}$ symmetry are easily identified by their appearance in the *xx* (*yy*,*zz*) or *xz* spectra, respectively. The lines of $B_{1g}$ and $B_{3g}$ symmetry are allowed, but cannot be distinguished, in the *z'y* = *xy*/ *zy* spectra. The frequencies (in cm$^{-1}$) of all experimentally observed $A_g$ and $B_{2g}$ Raman lines are listed in Table I along with the mode frequencies predicted by LDC. For these calculation a shell model described in detail in Ref.[12] was used. This model gives an adequate description of the vibrations in perovskitelike structures because it accounts for their predominant ionicity. The ionic interactions are represented by long-range Coulomb potentials and short-range repulsive potentials of the Born-Mayer form *a*exp(-*br*) where *a* and *b* are constants and *r* is the interionic separation. The deformation of the electron charge den-sity of the ions is described in the dipole approximation considering each atom as consisting of a point charged core and a concentric spherical massless shell with charge $Y$. Each core and its shell are coupled together with a force constant $k$ giving rise to the free ionic polarizability $\alpha = Y^2/k$. The model parameters for the calcium, iron, and oxygen ions and their interaction potentials are taken from a previous study of simpler compounds with perovskitelike structure [12].

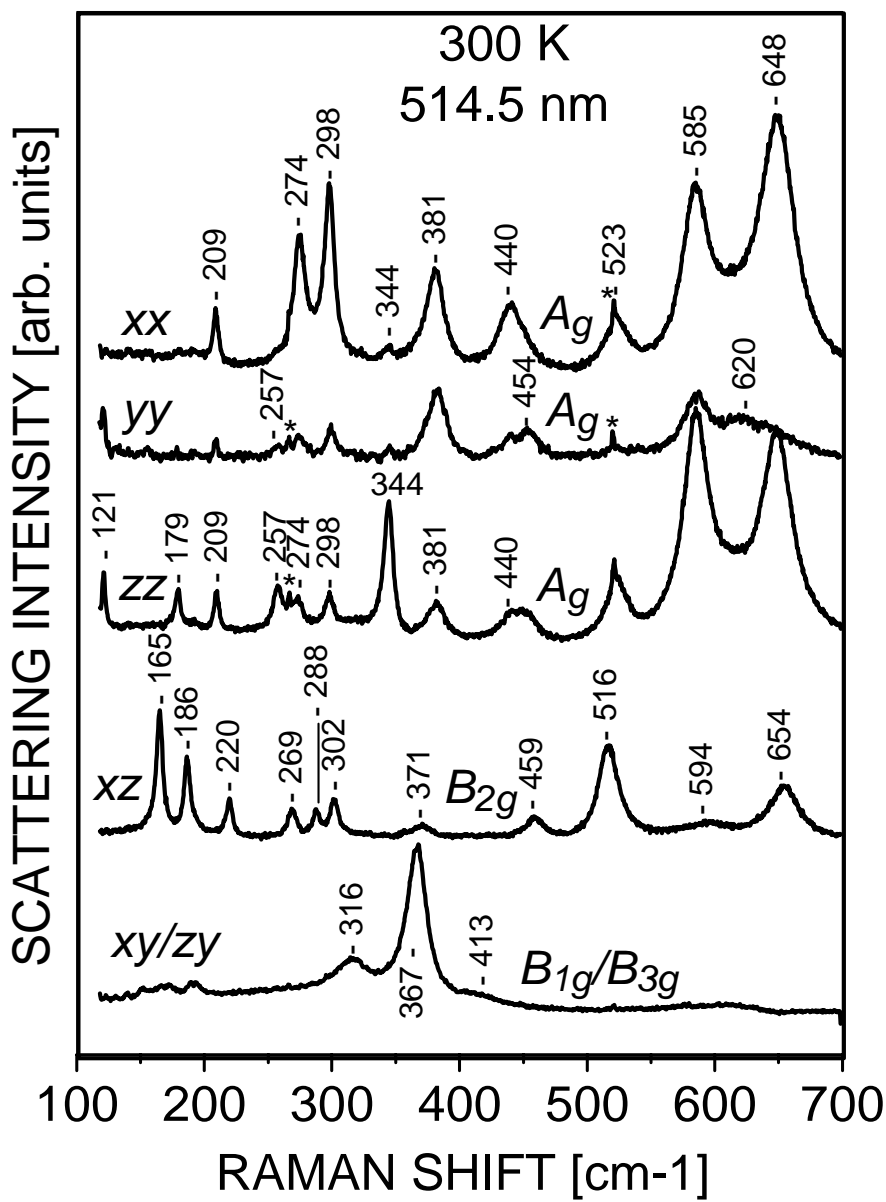

Fig.2 Raman spectra of $CaFe_2O_4$ obtained at room temperature with several exact scattering configurations.

Table I
Comparison of experimentally observed and theoretically predicted frequencies (in cm$^{-1}$) for the $A_g$ and $B_{2g}$ modes in CaFe$_2$O$_4$.

| Mode | Experiment 300 K | Experiment 20 K | Theory LDC | Main atomic motions (descending amplitudes) |
|---|---|---|---|---|
| $A_g(1)$ | 121? | | 84 | Fe1, Fe2, O3, O1, Ca |
| | 179? | 186 | | |
| $A_g(2)$ | 209 | 215? | 210 | Ca |
| $A_g(3)$ | 257 | 260 | 249 | Ca, Fe2, Fe1 |
| $A_g(4)$ | 274 | 280 | 267 | O1, O3 |
| $A_g(5)$ | 298 | 304 | 290 | O3, Fe1, Fe2 |
| $A_g(6)$ | 344 | 350 | 320 | O1, O3, Fe2, Fe1, Ca |
| $A_g(7)$ | 381 | 395 | 406 | O3, Ca, Fe2, Fe1 |
| $A_g(8)$ | 440 | 446 | 428 | O1, O4, Fe1, Fe2 |
| $A_g(9)$ | 454 | 465? | 451 | O2, O4, O1, O3 |
| $A_g(10)$ | | | 485 | O2, O4, O1, Fe1 |
| $A_g(11)$ | 523 | 531 | 523 | O2, O1, O4 |
| $A_g(12)$ | | | 547 | O2, O4 |
| $A_g(13)$ | 585 | 591 | 578 | O4, O1, O3, O2 |
| $A_g(14)$ | 648 | 655 | 653 | O3, O1 |
| | | | | |
| $B_{2g}(1)$ | 165 | 170 | 155 | O3, Fe2, O1 |
| $B_{2g}(2)$ | 186 | 190 | 183 | Fe1, Ca, Fe2, O4, O1 |
| $B_{2g}(3)$ | 220 | 223 | 224 | Ca, O1, O3 |
| | | 269? | | |
| | | 288? | | |
| $B_{2g}(4)$ | 302 | 305 | 298 | O3, Ca |
| $B_{2g}(5)$ | | | 325 | Ca, Fe1, O1 |
| $B_{2g}(6)$ | | | 337 | O1, Fe2, Ca |
| $B_{2g}(7)$ | 371? | | 407 | Fe1, Fe2, O2, O1 |
| $B_{2g}(8)$ | | 419 | 431 | O4, O3, O1 |
| $B_{2g}(9)$ | 459? | | 478 | O2, O1, O4 |
| $B_{2g}(10)$ | 516 | 523 | 516 | O2, O4, O3, O1 |
| $B_{2g}(11)$ | | | 536 | O2, O4, O1 |
| $B_{2g}(12)$ | | | 588 | O4, O3, O2, O1 |
| $B_{2g}(13)$ | 594? | 595 | 620 | O1, O2, O3 |
| $B_{2g}(14)$ | 654 | 661 | 663 | O3 |

Except for frequencies, the lattice dynamical calculations predict also the eigenvectors of the Γ-point phonon modes. Examples of atomic motions involved in some $A_g$ and $B_{2g}$ modes are shown in Fig.3. As a rule the motions are mixed and it is difficult to label the modes as "rotational", "bending" or "stretching". The last column of Table I show in descending order the atoms with most substantial contribution to a given mode.

Fig.3 Main atomic motions in some $A_g$ and $B_{2g}$ modes.

The variations with temperature of the position of most intensive $A_g$ lines are shown in Fig.4. With decreasing temperature all lines exhibit hardening commonly assigned to anharmonic scattering [13]. Within experimental error no detectable anomalies are observed near $T_N$.

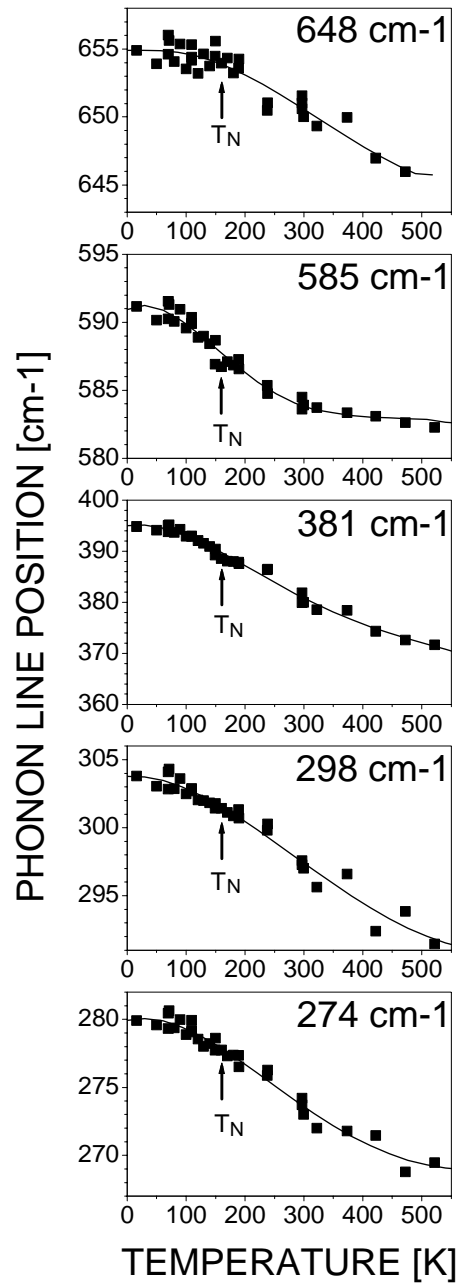

Fig.4 Variation with temperature of the position of most pronounced $A_g$ Raman lines.

In conclusion, the polarized Raman spectra from $CaFe_2O_4$ single crystals were measured and the symmetry of all observed phonon lines was determined. By comparison to the calculations of lattice dynamics, most of the lines were assigned to definite phonon modes. The temperature behavior of the Raman lines does not provide evidence for phonon anomalies near the magnetic ordering temperature.

This work was supported in part by the State of Texas through the Texas Center for Superconductivity and Advanced Materials (TcSAM) at the University of Houston and by the Bulgarian Science Fund (Project F-1207).